\documentstyle[aps,epsf,prd]{revtex}
\tightenlines
\begin{document}

\preprint{
\hfill$\vcenter{\hbox{\bf IFUSP/PRE-PRINT: 1320/98} 
                \hbox{\bf IFT-P.072/98} 
             }$ }

\title{ $CP$ VIOLATION IN VACUUM NEUTRINO OSCILLATION EXPERIMENTS}
\draft
\author{A.\ M.\ Gago$^{1,2}$ \thanks{Email: agago@charme.if.usp.br}, 
V.\ Pleitez$^3$\thanks{Email: vicente@ift.unesp.br},
and R.\ Zukanovich Funchal$^1$\thanks{Email: zukanov@charme.if.usp.br} }

\address{\em $^1$ Instituto de F\'{\i}sica, Universidade de S\~ao Paulo, \\
    C.\ P.\ 66.318, 05389-970 S\~ao Paulo, Brazil \\
      $^2$ Secci\'on F\'{\i}sica, Departamento de Ciencias, Pontificia 
       Universidad Cat\'olica del Per\'u \\
       AP 1762 Lima, Per\'u \\
    $^3$ Instituto de F\' {\i}sica Te\'orica -- UNESP \\
    R. Pamplona 145, 01405-900 S\~ao Paulo, Brazil. }

\maketitle

\vspace{.2in}


\hfuzz=25pt

\begin{abstract}

We discuss the use of the $CP$ asymmetry parameter ($A_{\scriptsize CP}$) 
as a possible observable of $CP$ violation in the leptonic sector.
In order to do this, we study for a wide range of values of $L/E$ 
the behavior of this asymmetry for the corresponding maximal  
value of the $CP$ violation factor allowed by all the present 
experimental limits on neutrino oscillations in vacuum and the recent 
 Super-Kamiokande atmospheric neutrino result. We work in the three neutrino 
flavor framework.

\end{abstract}



\section{Introduction}
\label{intro}

We are definitively  living in very exciting times in 
neutrino physics. 
The recent results from Super-Kamiokande (SK) 
indicating  evidence for neutrino oscillations 
in atmospheric showers~\cite{sk}, 
the reports of observed neutrino oscillations  by 
the Los Alamos Liquid Scintillator Detector (LSND)~\cite{lsnd1} in 
the $\overline{\nu}_\mu \rightarrow \overline{\nu}_e$ 
and  $\nu_\mu \rightarrow \nu_e$ channels 
in conjunction with other hints such as the 
results of solar neutrino experiments~\cite{homestake,gallex,sage,kamiokande} 
makes it difficult to believe today that all of these facts are not 
related to neutrino properties beyond the standard model.  

In the past neutrino physics has led to the discovery of neutral 
currents and provided the first indications in favor of the standard model 
of electroweak interactions. It may as well, if neutrino oscillations 
turn out to be confirmed by future experiments, 
reveal itself as an invaluable tool to cast some light on  physics
beyond the standard model, in particular, on the origin 
of $CP$ violation. One can hope this can be achieved in the study of the  
neutrino oscillation phenomena in some of the future 
experiments~\cite{k2k,minos,icarus}. 

The recent KTeV result on $Re(\epsilon'/\epsilon)$~\cite{ktev} finally 
 establishes direct $CP$ violation in the Kaon system, rules out once
and for all pure superweak theory and supports the notion of a nonzero
phase in the CKM matrix. This makes it even more interesting to check
if a similar effect also happens in the lepton sector.

As it is well known~\cite{cabibbo,barger,bilenky,pakvasa} 
$CP$ violation in neutrino oscillations can, in principle, be observed in 
neutrino experiments by looking at the differences of the transition 
probabilities between $CP$--conjugate channels, 
$\Delta  P  = P(\overline \nu_\alpha \rightarrow \overline \nu_\beta)-P(\nu_\alpha \rightarrow \nu_\beta)$. 
It has been pointed out by many authors that it may be, in practice,
very difficult to get a reliable measurement  of $\Delta  P$ 
due to possible Earth matter effects in long baseline neutrino 
experiments~\cite{matter,nuno}.
However we believe it is worthwhile to try to evaluate the
maximal size of $CP$ violation effect in vacuum since 
the vacuum oscillation experiments will be the ultimate  proof 
of electroweakly induced neutrino flavor conversion. 
Both solar and atmospheric neutrino data can,  in fact,  be explained by 
alternative mechanisms invoking  matter phenomena~\cite{guzzo,kb,vic}.

Here we will use the asymmetry parameter,
$A_{\scriptsize CP} = \Delta  P/ [ P ( \nu_\alpha \rightarrow \nu_\beta 
)+ P( \overline \nu_\alpha \rightarrow \overline \nu_\beta) ] $,
suggested by Cabibbo~\cite{cabibbo},  as an alternative to measure
$CP$ violation in the leptonic sector. We will investigate the possible
values of this parameter and the corresponding maximal values of  
$\Delta  P$  allowed by present experimental data for different  
$L/E$ situations, for a particular choice of neutrino mass squared 
differences.  We will not make any assumption about the elements of
the mixing matrix.

\section{Scale of masses and CP violation parameters}
\label{sec1}

This work will be developed in the three flavor neutrino scheme 
and for this reason only two mass scale indications can be taken to be 
right~\cite{smirnov}. We will fix them to be: 
\begin{equation}
       \Delta  m^{2}_{21}\approx 3.0 \times 10^{-3}~\mbox{eV}^{2}, \quad
       \Delta  m^{2}_{32}\approx 0.4~\mbox{eV}^{2},
\label{masses} 
\end{equation}

\noindent which are taken within the allowed regions given by the 
atmospheric~\cite{sk} and terrestrial neutrino   
experiments~\cite{lsnd1,bugey,e776} respectively. 
We will also take into account the probability constraints 
coming from this two types of experiments.
We admit here that the solar neutrino problem may be understood
invoking other types of mechanisms~\cite{guzzo,mansour}.

We can get the analytical expressions for $\Delta  P$ and $A_{\scriptsize CP}$ 
using the usual form of the CKM matrix parameterization:

\begin{equation}
\text{U}=\left[
\begin{array}{ccc}
c_{12}c_{13} & s_{12}c_{13} & s_{13}e^{-i\delta } \\
-s_{13}c_{23}-c_{12}s_{23}s_{13}e^{i\delta } &
c_{12}c_{23}-s_{12}s_{23}s_{13}e^{i\delta } & s_{23}c_{13} \\
s_{12}s_{23-}c_{12}c_{23}s_{13}e^{i\delta } &
-c_{12}s_{23-}s_{12}c_{23}s_{13}e^{i\delta _{}} & c_{23}c_{13}
\end{array}
\right], \label{km}
\end{equation}  
where $c$ and $s$ denote the cosine and the sine of the  
respective arguments.

Thus  $\Delta  P$ in vacuum can be written as:
\begin{equation}                                                             
\Delta  P(\alpha,\beta)=P(\overline{\nu}_{\alpha} \rightarrow \overline{\nu}_{\beta})-P(\nu_{\alpha} \rightarrow \nu_{\beta})=4 J_{\scriptsize CP}(\sin \triangle_{12}+\sin \triangle_{23}+\sin \triangle_{31}),
\label{dP}
\end{equation}
with  $\alpha , \beta= e,\mu,\tau$ and 
\begin{equation}
\triangle_{ij}= 2.54 \left ( \frac {\Delta  m^{2}_{ij}}{1 \mbox{eV}^2} \right ) \left ( \frac {L}{\mbox{km}} \right ) \left ( \frac {1 \mbox{GeV}}{E} \right ), ~ i,j=1,2,3;
\label{tri}
\end{equation}
where $\Delta  m^{2}_{ij}=m^{2}_{j}-m^{2}_{i}$ and the well known Jarlskog invariant~\cite{ceci}
\begin{equation}
J_{\scriptsize CP}=c^{2}_{13}s_{13}c_{12}s_{12}c_{23}s_{23}\sin \delta.
\label{jcp}
\end{equation}

We can see from Eqs.\ (\ref{dP}-\ref{jcp}) that in vacuum  
$\Delta P(\mu,e)=\Delta P(\mu,\tau)= \Delta P(e,\tau )$ so they are all simply 
referred to as $\Delta P$.

On the other hand $A(\alpha,\beta)_{\scriptsize CP}$, which depends on 
the specific channel $(\alpha,\beta)$, is given by :
\begin{equation}
A(\alpha,\beta)_{\scriptsize CP} = \frac{P(\overline{\nu}_{\alpha} \rightarrow \overline{\nu}_{\beta})-P(\nu_{\alpha} \rightarrow \nu_{\beta})}{P(\overline{\nu}_{\alpha} \rightarrow \overline{\nu}_{\beta})+P(\nu_{\alpha} \rightarrow \nu_{\beta})}.
\label{acp}
\end{equation}

In practice, for a real experiment, one has to  average  the probabilities 
in Eqs.\ (\ref{dP}) and (\ref{acp}) over the corresponding
experimental conditions, i.e. neutrino energy spectrum, distance $L$, 
efficiencies etc. 

\section{Experimental Constraints}
\label{sec2}

We have studied the experimental constraints imposed on the neutrino 
oscillation probabilities by several experimental results. In order 
to  obtain realistic  constraints we have, for each experiment, 
averaged the corresponding probability over its  distributions of $L$ and  
neutrino energy. This will be denoted  by  $\langle \cdots \rangle$ 
in what follows.

In the terrestrial case  we have taken the results from two short baseline 
accelerator experiments LSND~\cite{lsnd1} and E776~\cite{e776} for the 
channels $\overline{\nu }_\mu \rightarrow \overline{\nu }_e$ and 
$\nu _\mu \rightarrow \nu_e$ respectively. To analyze  LSND  we have
used the simple model quoted in Ref.~\cite{pdg}. In the case of E776 the
neutrino beam energy spectrum was taken from Ref.~\cite{e776}. \\

\noindent LSND: 
\begin{equation} 
\langle P(\overline{\nu }_\mu \rightarrow \overline{\nu}_e) \rangle=0.31\pm 0.10\pm 0.05 \%  
 ; \quad L \sim 30\mbox{ m}, \quad E\sim 36-60 \,\, \mbox{MeV},
\label{plsnd}
\end{equation} 

\noindent E776:
\begin{equation}
\langle P({\nu }_\mu \rightarrow {\nu}_e) \rangle \leq 1.5 \times 10^{-3} \, \mbox{ at } 90 \% 
 \, \mbox{C.L.}; \quad   
L/E \sim 1.
\label{p776}
\end{equation}

For the oscillation channel $\nu _\mu \rightarrow \nu_\tau$ we have used 
the data from CHORUS~\cite{chornom}. 
The  CHORUS relevant information is available at their Web 
site ~\cite{chorus_page}. This 
short baseline accelerator experiment gives 

\begin{equation}
\langle P({\nu }_\mu \rightarrow {\nu}_{\tau}) \rangle \leq 6 \times 10^{-4} \, \mbox{ at } 90 \% \, \mbox{C.L.}; \quad
L/E\sim 0.02.
\label{pchor}
\end{equation}

We have obtained from CHOOZ~\cite{chooz}, a long baseline reactor  
experiment searching for the disappearance of the 
$\overline{\nu}_e$, the limit :

\begin{equation}
1-\langle P(\overline{\nu }_e \rightarrow \overline{\nu}_e) \rangle 
\leq 10^{-1} \, \mbox{ at } 90 \% 
 \, \mbox{C.L.}; \quad   
L/E\sim 300 \, ,
\label{pchooz}
\end{equation}

\noindent and from the Bugey~\cite{bugey} reactor experiment 

\begin{equation}
\begin{array}{l}
1- \langle P(\overline{\nu }_e \rightarrow \overline{\nu}_e) \rangle 
\leq 10^{-2} \, \mbox { at } 90 \% 
 \, \mbox{C.L.}; \quad   
L=15,40,95 \, \mbox{m}, \quad E \sim 1-6 \,\, \mbox{MeV}.
\end{array}
\label{pbugey}
\end{equation}

We have read out the positron energy spectrum for CHOOZ from Ref.~\cite{chooz}
and for Bugey from Ref.~\cite{bugey} and used the relation 
$E_{\nu}=E_{e^+}+1.8 \mbox{ MeV}$ to extract  the neutrino energy spectrum. 

\begin{figure}[ht]
\centering\leavevmode
\epsfxsize=250pt
\epsfbox{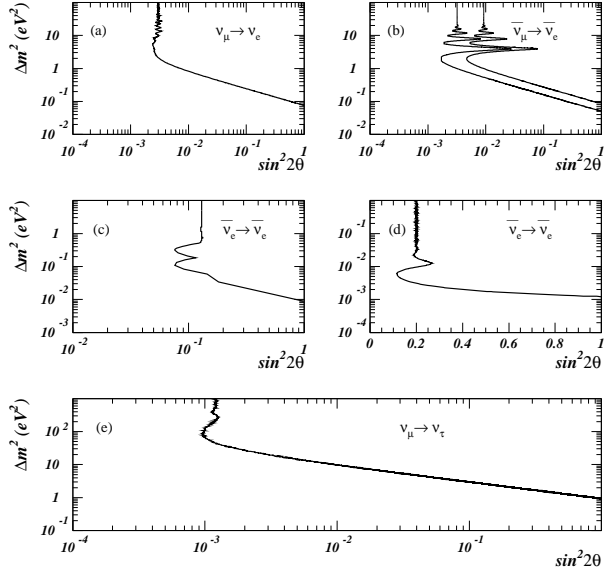}
\vglue -0.01cm
\caption{Our reproduction of the exclusion/allowed regions for different 
experiments in two generations: (a) E776, (b) LSND, (c) Bugey, (d)
CHOOZ and (e) CHORUS. The exclusion regions at 90\% C.L. are the ones
to the right of the curves for all experiments except LSND. In the
case of LSND the allowed region is the region between the two curves.} 
\label{expt}
\end{figure}

We show in  Fig.~\ref{expt}  our own reproduction of the
exclusion (allowed in the case of LSND) regions of these experiments. 
We observe a reasonable agreement with
the original experimental plots that can be found in 
Refs.~\cite{lsnd1,bugey,e776,chornom,chooz}. This is a demonstration
that  Eqs.\ (\ref{plsnd}-\ref{pbugey})  correctly represent  the constraints 
coming from these reactor and accelerator neutrino  experiments. 

In the case of atmospheric neutrinos, we only have  
used the latest SK~\cite{sk} result since it is the most 
precise one. We have performed our analysis of SK atmospheric 
data in the following way. 
We have separately used the muon neutrino events
($\mu$-like events) ratio 

\begin{equation}
R_\mu(L/E)= \gamma \left( \langle P_{\mu\mu}(L/E) \rangle +\frac{e_0}{\mu_0}
\langle P_{e\mu}(L/E) \rangle \right),
\label{rmu}
\end{equation} 
 
and the electron neutrino events ($e$-like events) ratio 

\begin{equation}
R_e(L/E)= \gamma \left( \langle P_{ee}(L/E) \rangle +\frac{\mu_0}{e_0}
\langle P_{\mu e}(L/E) \rangle \right),
\label{re}
\end{equation} 
where $\mu_0=\mu_0(L/E)$ and $e_0=e_0(L/E)$ are the distributions of Monte
Carlo $\mu$-like and $e$-like events  taken from Ref.~\cite{thesis}
 and $\gamma$ is a  constant to take into account  the overall 
normalization. 
We have taken the value of this constant to be 1.16 in 
accordance with Ref.\ \cite{bb}. We have introduced in Eqs.\ (\ref{rmu}) and
(\ref{re}) the shorthand  $P_{\alpha \beta}= P(\nu_\alpha \to
\nu_\beta), \alpha,\beta=e,\mu$. 
The probabilities $\langle P \rangle$ have been smeared by the
resolution function given in Ref.\ \cite{JGL} at each $L/E$ value. 

 On the other hand, from Fig.\ 4 of Ref.~\cite{sk} 
we can read eight different values of $L/E$ and their 
corresponding $\mu$-like and $e$-like ratios.
Using this we have calculated the simple average of each ratio
obtaining: 
 
\begin{equation}
\overline R_\mu  = 0.76 \pm 0.08; \, \,  \overline R_e = 1.19 \pm 0.13,
\label{remu}
\end{equation}
defining two allowed bands, one  for the $e$-like events  and 
the other for the $\mu$-like events ratio.
These are our selection criteria coming from SK data which were used
in the following way: for a given  set of 
values of the mass squared differences, the mixing angles and the phase 
we have computed the right hand side of Eq.\ (\ref{rmu}) and
Eq.\ (\ref{re}) taking an average over the eight $L/E$ bins and verifying 
if they satisfy the conditions given by Eqs.\ (\ref{remu}).
It is important to stress that working with $\mu$-like and $e$-like
event ratios separately has the advantage that we are able to 
constraint independently $P(\nu_\mu \rightarrow \nu_e)$ and 
$P(\nu_e \rightarrow \nu_\mu)$. Indeed doing this  we observe that
these probabilities are  always below a few \%, this is 
much more stringent that the SK analysis shown in Ref.\ \cite{thun,olsson}.

It must be noted that this simple SK analysis is not totally rigorous,
in fact by using the $L/E$ distribution from Ref. \cite{sk} 
we only work with a sub-sample of the SK data 
(fully-contained events). Moreover there are quite large uncertainties 
in the determination of this distribution due to the fact that one
relies on the observed final lepton to infer the neutrino physical quantities,
these uncertainties are discussed in Ref. \cite{thesis}. 
Nevertheless we consider our approach good  enough for the goals of this paper.

Since the Eqs.\ (\ref{plsnd}-\ref{remu})  
are in fact independent of the number of 
neutrino generations we are now free to use these probability limits in the 
three neutrino flavor framework.

\section{Limits of J$_{CP}$, $\Delta  P$ and $A_{ \scriptsize CP} $ }
\label{sec3}

In order to determine the maximum permitted values 
of  $\Delta  P$,  $A(\mu,e)_{\scriptsize CP}$ and 
$A(\mu,\tau)_{\scriptsize CP}$ we proceed in 
the following way. We choose randomly different values of
the mixing parameters $s^{2}_{12},s^{2}_{23}$ and $s^{2}_{13}$ taken in the
interval $[0,1]$. For each drawn set of these parameters  we 
evaluate the values corresponding to  Eqs.\ (\ref{plsnd}-\ref{re}).
We check if  they simultaneously pass  all the experimental
constraints given in  Eqs.\ (\ref{plsnd}-\ref{pbugey}) 
and Eq.\ (\ref{remu}). It really means that we look for a 
common  allowed region among all experiments. 
In the positive case we compute the $CP$ violation factor
$J_{\scriptsize CP}$ for fixed values of the phase angle $\delta$.
We select among them the maximal value of $J_{\scriptsize CP}$, 
$J_{\scriptsize CP}^{\mbox{\scriptsize max}}$, for each $\delta$ and
calculate the corresponding  values of $\Delta  P^{\mbox{\scriptsize max}}$,  
$A(\mu,e)_{\scriptsize CP}$ and $A(\mu,\tau)_{\scriptsize CP}$ 
for $L/E$ varying from 40 to 250.

In Tables~\ref{jmax1}-\ref{jmax2} we show  
$J_{\scriptsize CP}^{\mbox{\scriptsize max}}$ 
for $\Delta m^2_{21} \approx 3.0 \times 10^{-3}$ eV$^2$ and 
$\Delta m^2_{32} \approx 0.27,2.0$ eV$^2$. We have picked these values 
so one can have an idea of the order of the  variations that 
$J_{\scriptsize CP}^{\mbox{\scriptsize max}}$ will have if one 
varies $\Delta m^2_{32}$ inside the LSND allowed region.
In general we see that the values of 
$J_{\scriptsize CP}^{\mbox{\scriptsize max}}$ in the squared mass 
difference interval considered in this paper  are more or less stable 
and  of ${\cal O}(10^{-3})$ in agreement with Ref.\ \cite{bilenky2}.

\begin{figure}[ht]
\centering\leavevmode
\epsfxsize=230pt
\epsfbox{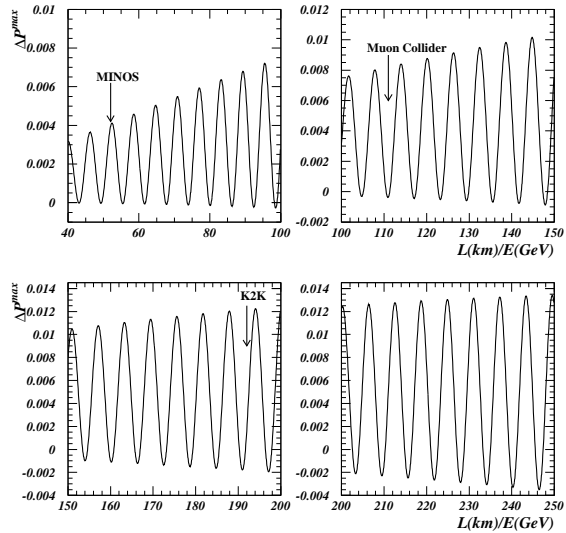}
\vglue -0.01cm
\caption{$\Delta P^{\mbox{\scriptsize max}}$ as a function of
$L/E$ for $\sin \delta$ = 1. }
\label{deltap}
\end{figure}

In Fig.~\ref{deltap} we show  
$\Delta  P^{\mbox{\scriptsize max}}$ as a 
function of $L/E$ for $\sin\delta$ = 1.
As one could guess the oscillating behavior of this curve is dictated
by the sum of sines in Eq.\ (\ref{dP}).
We are not showing here  other curves for different values of $\sin\delta$
since they have exactly the same form as these ones, only the
oscillation amplitude will change in accordance to the corresponding 
$J_{\scriptsize CP}^{\mbox{\scriptsize max}}$.
We observe that the maximal value of $\Delta  P^{\mbox{\scriptsize
max}}$, which increase with $L/E$, are of the  ${\cal O}(10^{-2})$. This 
is in accordance with the estimations given in Refs. ~\cite{nuno,bilenky2}.

\begin{figure}[ht]
\centering\leavevmode
\epsfxsize=220pt
\epsfbox{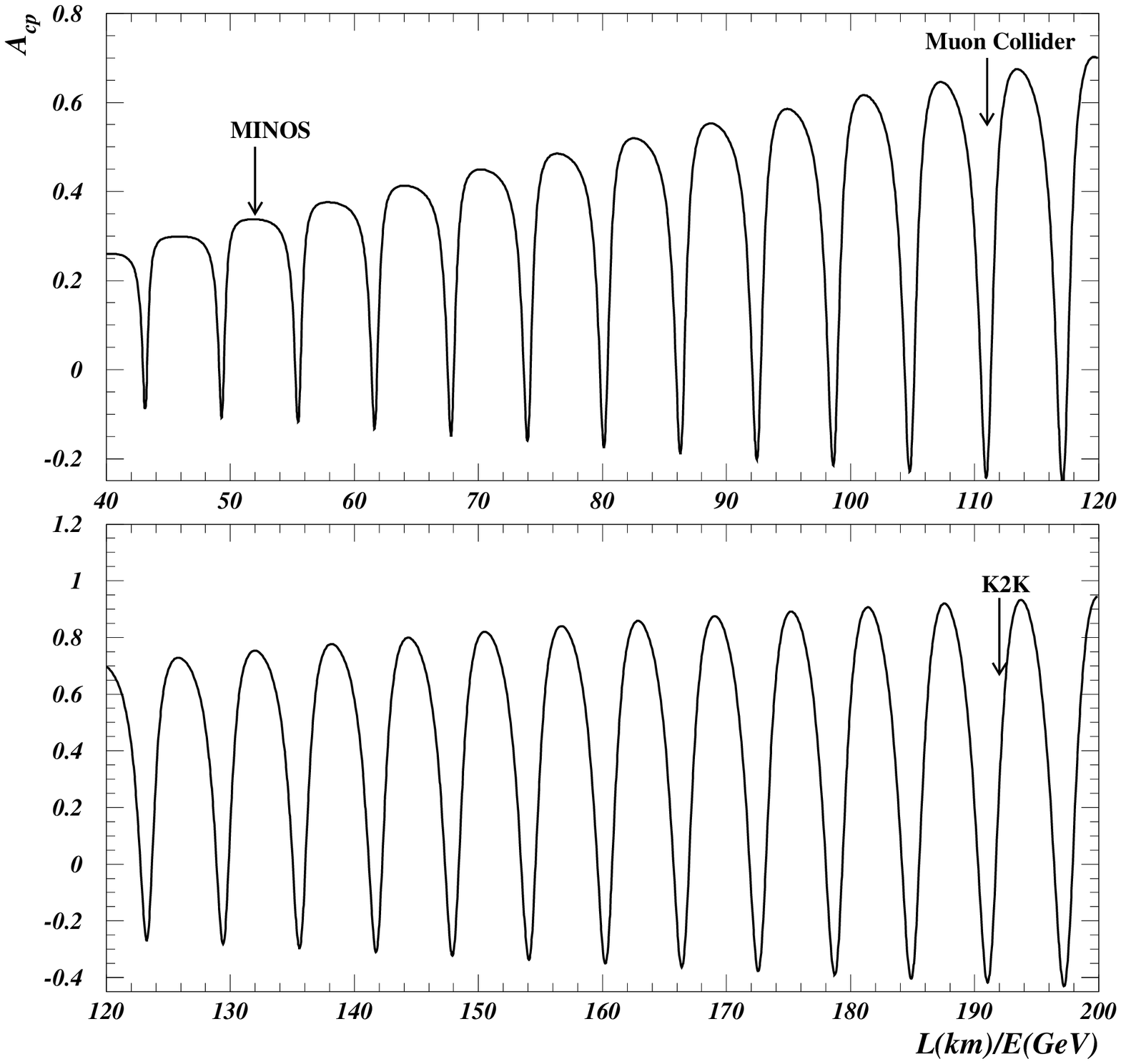}
\vglue -0.01cm
\caption{ Values of $A(\mu,e)_{\mbox{$CP$ }}$   as a 
function of $L/E$ for 
$\sin \delta$ = 1. }
\protect \label{acpfig1}
\end{figure}

\begin{figure}[ht]
\centering\leavevmode
\epsfxsize=220pt
\epsfbox{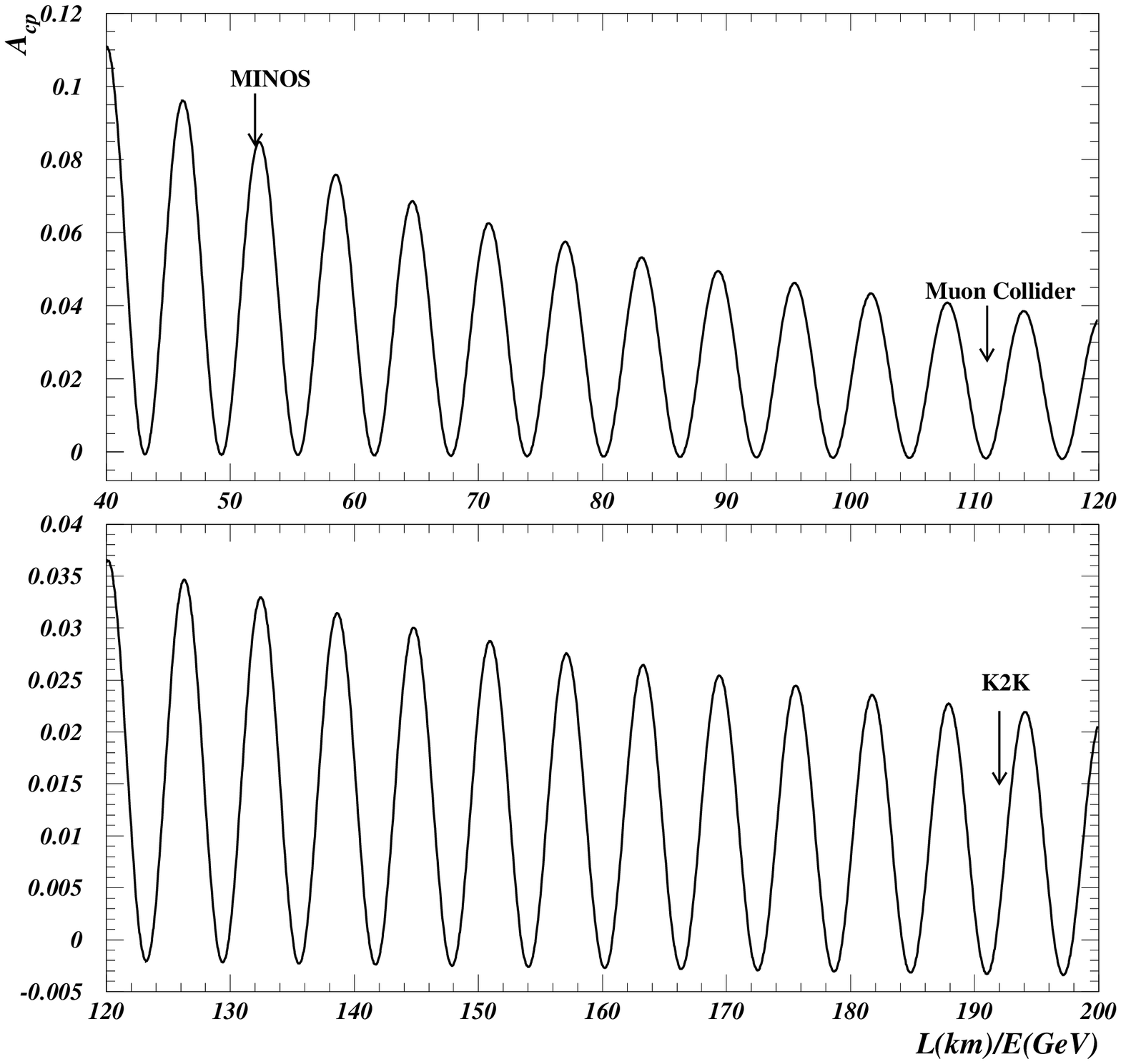}
\vglue -0.01cm
\caption{ Values $A(\mu,\tau)_{\mbox{$CP$ }}$    as a 
function of $L/E$ for 
$\sin \delta$ = 1. }
\protect \label{acpfig2}
\end{figure}

In  Figs.~\ref{acpfig1}-\ref{acpfig2}  we show the behavior of  
$A(\mu,e)_{\scriptsize CP}$ and $A(\mu,\tau)_{\scriptsize CP}$   
as a function of $L/E$ for $\sin\delta$=1.
We can observe that 
$A(\mu, e)_{\scriptsize CP}$ grows and 
$A(\mu, \tau)_{\scriptsize CP}$ decreases as  a function of 
 $L/E$.  Also the maximal values for $A(\mu, e)_{\scriptsize CP}$ 
are of the  ${\cal O}(1)$ while for $A(\mu, \tau)_{\scriptsize CP}$
they are of the ${\cal O}(10^{-1})$. 
This result was expected since we know that the only difference 
between $A(\mu, e)_{\scriptsize CP}$ and  $A(\mu, \tau)_{\scriptsize CP}$,
in vacuum, comes from the denominator of Eq.\ (\ref{acp}) and 
$P(\nu_\mu \rightarrow \nu_e)$ is currently very 
much suppressed by data while SK data seems to support 
$P(\nu_\mu \rightarrow \nu_\tau)$ oscillations that can reach the order of 
$10^{-1}$ for neutrinos coming from bellow the horizon.

As we have already mentioned at the end of Sec. \ref{sec1} one has to
be careful in interpreting our results in relation to future
experiments and remember to take into account 
the average over the corresponding $L/E$
distribution.  Because we did not want to make any ad-hoc hypothesis 
on $L/E$ distribution  we have deliberately chosen not to do any smearing on 
$L/E$. We believe that presenting our results in this unfolded way make
them more useful and ready to be applied to any real experimental situation.

To illustrate the implications of our results in future  experiments
we have put an arrow in Fig.~\ref{deltap}-\ref{acpfig2} at the corresponding 
 mean value of $L/E$ for MINOS~\cite{minos}, K2K~\cite{k2k} and 
one of the possible configurations of a long baseline neutrino 
experiment using neutrinos from  a muon collider~\cite{muonc}.   
Based on our results we also have computed an  estimation, 
averaging over the expected energy spectrum, of the maximal values
that can be investigated at MINOS:  
$\langle A(\mu,e)_{\scriptsize CP} \rangle = 0.33$ ,
$\langle A(\mu,\tau)_{\scriptsize CP} \rangle = 0.02$,
$\langle \Delta P\rangle = 0.0022$; 
and K2K:
 $\langle A(\mu,e)_{\scriptsize CP} \rangle = 0.19$ ,
$\langle A(\mu,\tau)_{\scriptsize CP} \rangle = 0.0017$,
$\langle \Delta P\rangle = 0.0014$. 
All these values are, as expected,  lower than the ones shown by the 
corresponding arrows in Fig.~\ref{deltap}-\ref{acpfig2} since 
in the $L/E$ scope of MINOS/K2K the dominant scale is 
$\Delta m^2_{21}$, the contribution from $\Delta m^2_{32}$ being
averaged out.

\section{Conclusions}
\label{conc}

We have found the maximal allowed values 
for $\Delta P^{\mbox{\scriptsize max}}$,
$A(\mu, e)_{\scriptsize CP}$ and $A(\mu, \tau)_{\scriptsize CP}$
as a function of $L/E$ and $\sin\delta$.
This was done in the three neutrino flavor framework
using the most stringent constraints from recent neutrino data and
admitting the two mass squared differences  to be 
$\Delta  m^{2}_{21}\approx 3.0 \times 10^{-3}~\mbox{eV}^{2}$ and
$\Delta  m^{2}_{32}\approx 0.4  ~\mbox{eV}^{2}$.
In fact this is in a way  complementary to Ref.\ \cite{tanimoto}.

It is important to remark that  we have adopted here a different approach
from the authors of Refs.\ ~\cite{nuno,bilenky2,tanimoto}  
 since we do not make any assumptions about the mixing parameters
other than the two squared mass  difference scales
and work with the  probability expression without any
approximation. Besides we have explicitly used the experimental 
resolution functions in our calculations.

We have seen that in general the values of $A_{\scriptsize CP}$ are much 
more sizable than the corresponding $\Delta P^{\mbox{\scriptsize max}}$. 
In fact, admitting the mass hierarchy considered in this paper, 
we have shown that the present  neutrino  data  tremendously  suppress 
the maximal values $\Delta P$ that can  be  investigated at the  next generation of neutrino experiments.  In addition 
since in the asymmetry the systematic errors cancel out, even if 
the absolute flux of the neutrino
beam is determined with an accuracy of 10 $\%$ it may 
be possible to measure $CP$
violation at 1 $\%$ level. This is 
particularly interesting since in long baseline
experiment there are expected matter effects that 
fake genuine $CP$ violation ~\cite{nuno}. 
In a forthcoming paper~\cite{gpz} we will  discuss the implications  
that the inclusion of matter effects will have on our present limits 
as well as how these limits change if one varies the squared 
mass difference scales.


\acknowledgements

This work was supported by Conselho Nacional de Desenvolvimento 
Cient\'{\i}fico e Tecnol\'ogico (CNPq), by Funda\c{c}\~ao de Amparo 
\`a Pesquisa do Estado de S\~ao Paulo (FAPESP), and by 
Programa de Apoio a N\'ucleos de Excel\^encia (PRONEX).



\begin{table}[htbp]
\begin{center}
\begin{tabular} {|c|c|c|}
$4 \times J_{\mbox{\scriptsize $CP$ }}^{\mbox{\scriptsize max}}$ & $4
\times J_{\mbox{\scriptsize $CP$ }}^{\mbox{\scriptsize max}}/
\sin \delta  $  & $\sin\delta$    \\
\hline
$0.0011$  & $0.0044$ & $0.2588$ \\
$0.0024$  & $0.0048$ & $0.5000$ \\
$0.0032$  & $0.0045$ & $0.7071$ \\
$0.0046$  & $0.0053$ & $0.8660$ \\
$0.0052$  & $0.0054$ & $0.9659$ \\
$0.0052$  & $0.0052$ & $1.0000$ \\
\end{tabular}
\caption{Maximal values of the Jarlskog factor  
obtained for different values of $\sin \delta$ with $\Delta
m^{2}_{21}\approx 3.0 \times 10^{-3}\mbox{ eV}^{2}$ and
$\Delta  m^{2}_{32}\approx 0.27\mbox{ eV}^{2}$.  }
\label{jmax1}
\end{center}
\end{table}

\begin{table}[htbp]
\begin{center}
\begin{tabular} {|c|c|c|}
$4 \times J_{\mbox{\scriptsize $CP$ }}^{\mbox{\scriptsize max}}$ & $4
\times J_{\mbox{\scriptsize $CP$ }}^{\mbox{\scriptsize max}} /
\sin \delta  $  & $\sin\delta$    \\
\hline
$0.0010$  & $0.0041$ & $0.2588$ \\
$0.0021$  & $0.0042$ & $0.5000$ \\
$0.0031$  & $0.0044$ & $0.7071$ \\
$0.0038$  & $0.0043$ & $0.8660$ \\
$0.0042$  & $0.0044$ & $0.9659$ \\
$0.0043$  & $0.0043$ & $1.0000$ \\
\end{tabular}
\caption{Maximal values of the Jarlskog factor  
obtained for different values of $\sin \delta$ with $\Delta
m^{2}_{21}\approx 3.0 \times 10^{-3}\mbox{ eV}^{2}$ and
$\Delta  m^{2}_{32}\approx 2.0\mbox{ eV}^{2}$.  }
\label{jmax2}
\end{center}
\end{table}


\end{document}